\documentclass[
    aps,
    prl,
    twocolumn,
    superscriptaddress,
    amsmath,amssymb
    ]{revtex4-2}

\usepackage{graphicx}
\usepackage{geometry}
\usepackage[utf8]{inputenc}
\usepackage[T1]{fontenc}
\usepackage{setspace}
\usepackage{amsmath, amssymb, amsthm}
\usepackage{bbm}
\usepackage{float}
\usepackage{microtype}

\usepackage{tikz}
\usepackage{mathtools}
\usepackage{enumerate}
\usepackage[hypertexnames=false]{hyperref}
\usepackage{xcolor} 
\usepackage{MnSymbol} 
\usepackage{bbm} 
\usepackage{orcidlink} 

\newcommand{\denop}{\mathcal{D}}
\newcommand{\hs}{\mathcal{H}}
\newcommand{\linop}{\mathcal{L}}
\DeclareMathOperator{\mytr}{Tr}
\newcommand{\id}{\mathbbm{1}}

\begin{document}

\title{Witnessing Polarization Entanglement of Annihilation Photons via Compton Scattering}

\author{Tobias C. Sutter\,\orcidlink{0009-0008-3179-7979}}
\email{tobias.christoph.sutter@univie.ac.at}
\affiliation{University of Vienna, Faculty of Physics, Währingerstrasse 17, 1090 Vienna, Austria.}
\author{Beatrix C. Hiesmayr\,\orcidlink{0000-0001-9062-6039}}
\email{beatrix.hiesmayr@univie.ac.at}
\affiliation{University of Vienna, Faculty of Physics, Währingerstrasse 17, 1090 Vienna, Austria.}
\affiliation{IT:U Interdisciplinary Transformation University, Freistädter Strasse 400, 4040 Linz, Austria.}

\date{\today}

\begin{abstract}
Experimentally certifying polarization entanglement in annihilation photons is challenging because polarizers are unavailable in the high-energy regime and one needs to rely on Compton polarimetry.
Previously proposed schemes designed for this task are either experimentally unfeasible or require a specific source state model.
We address these issues by reformulating Compton scattering as a continuous-outcome quantum instrument, thereby enabling the prediction of the scattering angles and the post-scattering polarization state.
Our framework allows connecting Compton polarimetry with entanglement witnesses by deriving experimentally accessible lower bounds on Bell state fidelities for arbitrary initial two-photon polarization states.
We discuss how this witness can be extracted from coarse-grained coincidence measurements with detectors of almost any angular resolution.
The presented scheme does not require knowledge of angle-independent empirical factors like detector efficiencies and source activity.
Thus, our results provide a feasible route towards experimental entanglement certification in para-positronium decay with current technology.
\end{abstract}

\maketitle

\textit{Introduction.}---
Quantum electrodynamics (QED) predicts that the photons created in para-positronium (para-Ps) decay have entangled polarizations, a property already anticipated in a pioneering study of electron-positron annihilation \cite{Wheeler_Polyelectrons_1946}.
However, experimentally establishing this prediction is difficult because the polarization of 511-keV photons cannot be probed using conventional polarization optics \cite{Bernard_GammaRayPolarimetry_2022}.
Instead, it is commonly accessed through Compton polarimetry \cite{DelMonte_ComptonPolarimetry_2023}, where the polarization dependence of the Compton cross section provides an experimentally accessible polarization measurement, albeit intrinsically unsharp and insensitive to circular polarization.
Early experiments respecting this limitation corroborated the theoretical predictions by measuring Compton scattering correlations \cite{Pryce_AngularCorrelationEffects_1947, Snyder_AngularCorrelationScattered_1948, Bleuler_CorrelationStatesPolarization_1948, Hanna_PolarizationAnnihilationRadiation_1948, Wu_AngularCorrelationScattered_1950, Bohm_DiscussionExperimentalProof_1957, Kasday_Angularcorrelationcomptonscattered_1975}.

Nowadays, renewed interest in certifying entanglement in annihilation photons stems from the potential to improve medical positron emission tomography (PET) imaging \cite{Kozuljevic_polarizationenhancedPETStudy_2026, Moulin_BenefitsQuantumEntanglement_2026, Watts_Photonquantumentanglement_2021, Moskal_Nonmaximalentanglementphotons_2025}.
Consequently, substantial recent effort has been dedicated to confirm entanglement experimentally \cite{Watts_Photonquantumentanglement_2021, Bordes_FirstDetailedStudy_2024,  Moskal_Nonmaximalentanglementphotons_2025, Ivashkin_Testingentanglementannihilation_2023, Bala_Probingarbitrarypolarized_2025}, accompanied by numerous theoretical advancements \cite{Hiesmayr_WitnessingEntanglementCompton_2019, Caradonna_StokesparameterrepresentationCompton_2024, Pei_Canmirrorsymmetry_2026, Caradonna_Comptonscatteringmediated_2025, Caradonna_Polarizationsymmetrysignature_2026, Caradonna_CoincidenceconditionedComptonmodulationrecovery_2026, Chen_QuantumchannelsCompton_2026, Hiesmayr_Quantumerrorchannels_2024, Joos_Entangledphotonsparapositronium_2026, Clarke_BellTestPhotons_2026}.
It is worth highlighting Refs.~\cite{Hiesmayr_Quantumerrorchannels_2024, Joos_Entangledphotonsparapositronium_2026, Clarke_BellTestPhotons_2026} as attempts to treat Compton scattering in a quantum information-theoretic framework, thereby simultaneously contributing to probing quantum information theory at high energies.

Crucially, however, even though Refs.~\cite{Watts_Photonquantumentanglement_2021, Bordes_FirstDetailedStudy_2024, Moskal_Nonmaximalentanglementphotons_2025, Caradonna_StokesparameterrepresentationCompton_2024, Hiesmayr_WitnessingEntanglementCompton_2019, Pei_Canmirrorsymmetry_2026} have established valuable angular-correlation results, Ref.~\cite{Clarke_BellTestPhotons_2026} rightfully remarks that the criteria, as formulated, do not witness entanglement unconditionally.
Most of these studies rely on additional assumptions about the photons' polarization to certify entanglement.
Otherwise, separable states can mimic the observed correlations, a problem that has already plagued Refs.~\cite{Pryce_AngularCorrelationEffects_1947, Snyder_AngularCorrelationScattered_1948, Bleuler_CorrelationStatesPolarization_1948, Hanna_PolarizationAnnihilationRadiation_1948, Wu_AngularCorrelationScattered_1950, Bohm_DiscussionExperimentalProof_1957, Kasday_Angularcorrelationcomptonscattered_1975}.
Indeed, the main challenge for entanglement certification is \textit{not} to show that the observed correlations are consistent with some entangled state, but that they are \textit{in}consistent with all separable states.
Mathematically, this corresponds to the concept of entanglement witnesses \cite{Chruscinski_Entanglementwitnessesconstruction_2014}.
Recent theoretical studies have addressed this issue \cite{Clarke_BellTestPhotons_2026, Joos_Entangledphotonsparapositronium_2026, Caradonna_Polarizationsymmetrysignature_2026}, although there is no clear path toward presently feasible experimental validation \cite{Supp_Mat}.
Thus, the question remains: Can finite-resolution Compton coincidence measurements rigorously certify the polarization entanglement of an otherwise unknown initial photon pair?

In this letter, we answer this question affirmatively.
To this end, we derive the continuous-outcome quantum instrument for Compton scattering from the underlying scattering dynamics.
We can thus compute the conditional outgoing photon polarization state and recover the scattering probabilities through the associated positive operator-valued measure (POVM).
This provides our foundation for deriving experimentally accessible polarization correlations without presupposing the initial polarization state.
We subsequently show how to determine a well-known entanglement witness from coarse-grained Compton measurements with almost arbitrary azimuthal and polar resolution.
The witness is a lower bound on the fidelity of the photons' polarization state and the maximally entangled Bell state predicted by QED for annihilation photons.
This represents a rigorous device-dependent entanglement certificate.
Furthermore, the presented scheme does not require knowledge of empirical, angle-independent factors like source activity or detector efficiency.
Hence, our results establish a concrete route towards quantitative entanglement certification in para-Ps decay using readily obtainable experimental observables.

\textit{From Stokes formalism to quantum map.}---
Consider a photon with polarization state $\rho$.
Its Stokes vector
\begin{align}\label{eq:stokes_vector}
    |S\rangle = \sum_{k=0}^{3} S_k |k\rangle
\end{align}
is determined by the expectation values $S_k =\langle \sigma_k \rangle_\rho = \mytr(\rho \, \sigma_k)$, where $\sigma_k$ are the Pauli matrices for $k=1,2,3$, and $\sigma_0=\id$.
Normalization of $\rho$ implies $S_0 = 1$.
Importantly, we associate the eigenvectors of $\sigma_x\equiv\sigma_1$, $\sigma_y \equiv\sigma_2$, and $\sigma_z \equiv \sigma_3$ with the polarization bases $\{|H\rangle,|V\rangle\}$, $\{|+45^\circ\rangle , |-45^\circ\rangle\}$, and $\{|R\rangle,|L\rangle\}$, respectively \cite{Supp_Mat}.
Hence, contrary to the convention in, e.g., Refs.~\cite{McMaster_MatrixRepresentationPolarization_1961, Hiesmayr_WitnessingEntanglementCompton_2019, Bala_Probingarbitrarypolarized_2025, Joos_Entangledphotonsparapositronium_2026}, we use the circular polarization basis as the computational basis.

Compton scattering of a polarized photon with energy $E_0$ \footnote{We measure energy in units of electron rest energy $m_e c^2$ such that $E_0 = 1$ corresponds to approximately $511\mathrm{keV}$.} on an unpolarized electron at rest, with the final electron spin remaining unobserved, can be described by transforming the Stokes vector as \cite{Fano_RemarksClassicalQuantumMechanical_1949, McMaster_MatrixRepresentationPolarization_1961}
\begin{align} \label{eq:stokes_trafo}
    |S\rangle \; \longmapsto \; |S'\rangle = T M |S\rangle \;.
\end{align}
Here, $T$ is the Fano transfer matrix and $M$ is the Müller rotation matrix given in \eqref{eq:T-matrix} and \eqref{eq:M-matrix} in the end matter, respectively.
The first entry of $|S'\rangle$ is the Klein-Nishina differential cross section per unit solid angle \cite{Klein_UeberStreuungStrahlung_1929}.

Although there exists a matrix $\rho'$ such that $S_i' = \mytr(\rho' \sigma_i)$, it cannot be interpreted as a post-scattering polarization state because its trace can exceed unity and has units of area per solid angle ($\mathrm{m}^2/\mathrm{sr}$).
We resolve this issue below, but for now treat the transformation of $\rho$ induced by \eqref{eq:stokes_trafo} as a linear map $\Phi: \rho \mapsto \Phi(\rho) =\rho'$ from a purely mathematical perspective.

To analyze $\Phi$ using tools from quantum information theory, we expand $\rho = \sum_{i,j=0}^{1} \rho_{i,j} |i\rangle\langle j|$ and $\rho' = \sum_{i,j=0}^{1} \rho'_{i,j} |i\rangle\langle j|$, where $|0\rangle = |L\rangle$ and $|1\rangle =|R\rangle$.
As discussed in the Supplementary Material \cite{Supp_Mat}, every linear map $\Phi$ has a so-called natural representation $F(\Phi)$, acting as $F(\Phi) |\rho\rrangle = |\Phi(\rho)\rrangle$ \footnote{Here, $|X\rrangle:= \sum_{i,j=0}^{1} x_{i,j} |i\rangle \otimes |j\rangle$ is the vectorization of the matrix $X = \sum_{i,j=0}^{1} x_{i,j} |i\rangle\langle j|$ (cf. Ref.~\cite{Watrous_TheoryQuantumInformation_2018}).}.
To determine $F(\Phi)$, we compute the invertible matrix $C$ such that $|S\rangle = C|\rho\rrangle$ and $|S'\rangle = C |\Phi(\rho)\rrangle$, leading to
\begin{align} \label{eq:def_Phi}
    C^{-1} TM C |\rho\rrangle = |\Phi(\rho)\rrangle \;.
\end{align}
Thus, the natural representation of $\Phi$ is given by $F(\Phi) = C^{-1} TM C$.
We give the explicit form of $C$ and $F(\Phi)$ in \eqref{eq:C-matrix}-\eqref{eq:nat_rep_of_Phi} in the end matter.
%
From this, it is straightforward to derive other representations of $\Phi$, and we present the Choi and Kraus representations in the end matter.
The former representation is useful for proving that $\Phi$ is completely positive (CP), i.e., it preserves the positivity of arbitrary quantum states; the latter is convenient for analyzing how Compton scattering transforms the photon's polarization.

\textit{Normalizing the linear Compton map}---
Now we resolve the issue that $\rho' = \Phi(\rho)$ is not a quantum state.
To obtain a physically meaningful transformation, consider the total cross section
\begin{align} \label{eq:sigma_tot_1}
    \sigma_\mathrm{tot}(\rho) = \int_{S^2} \mathrm{d}\Omega \;\mytr(\Phi(\rho)) \;,
\end{align}
where $S^2$ is the unit sphere in three dimensions.
For the standard case of Compton scattering from an unpolarized electron at rest, with final electron spin unobserved, rotational symmetry together with parity conservation for the scattering implies that the total cross section is independent of the incident photon polarization, i.e.,
\begin{align}\label{eq:sigma_tot_2}
    \sigma_\mathrm{tot}(\rho) = \sigma_\mathrm{KN} \mytr(\rho) \;.
\end{align}
Since $\sigma_\mathrm{KN}$ is positive and independent of $\rho$, we define
\begin{align} \label{eq:psi_instrument_density}
    \Psi := \frac{1}{\sigma_\mathrm{KN}} \Phi \;,
\end{align}
which is still CP and linear in $\rho$.
The trace of $\Psi(\rho)$ is
\begin{align} \label{eq:dpdOmega}
    \mytr(\Psi(\rho)) = \frac{1}{\sigma_\mathrm{KN}}\frac{\mathrm{d}\sigma_{\rho}}{\mathrm{d}\Omega} = \frac{\mathrm{d}p_\rho}{\mathrm{d}\Omega} \;,
\end{align}
which is a probability density per unit solid angle.
Thus, it is still not dimensionless and can exceed unity, as only $\frac{\mathrm{d}p_\rho}{\mathrm{d}\Omega} \geq 0$ and $\int_{S^2} \mathrm{d}\Omega \; \frac{\mathrm{d}p_\rho}{\mathrm{d}\Omega} = 1$ are required.
However, we can interpret $\Psi$ as the \textit{CP density of a quantum instrument} (with respect to $\mathrm{d}\Omega$).
To obtain a quantum instrument \cite{Davies_operationalapproachquantum_1970, Ozawa_Quantummeasuringprocesses_1984}, we define
\begin{align} \label{eq:CPTNI-map}
    \Psi_s := \int_s \mathrm{d}\Omega \; \Psi \;,
\end{align}
where $s\subseteq S^2$ corresponds to a finite angular detector size in the experiment.
This map satisfies
\begin{align}
    \mytr(\Psi_s(\rho)) = \int_s \mathrm{d} \Omega \; \frac{\mathrm{d}p_\rho}{\mathrm{d}\Omega} \;,
\end{align}
which is the probability that a photon with initial energy $E_0$ and polarization $\rho$ scatters into the solid angle interval $s\subseteq S^2$.
Therefore, $0\leq \mytr(\Psi_s(\rho)) \leq 1$, and thus $\Psi_s$ is CP and trace non-increasing.
Consequently, $\Psi_s(\rho)$ can be viewed as an unnormalized quantum state describing the post-scattering polarization state.
For $s=S^2$, the map $\Psi_{S^2}$ is CP and trace-preserving, i.e., a quantum channel. 
The discussion of this section is summarized in Table~\ref{tab:map_hierarchy}.
\begin{table}[t]
\centering
\begin{tabular}{| c | c | c|} 
     \hline
     \rule[-7.5pt]{0pt}{20pt}
     Map &  Interpretation & Problem \\
     \hline\hline
     \rule[-7.5pt]{0pt}{20pt}
     $\Phi$ &  CP cross-section density & units $\mathrm{m}^2/\mathrm{sr}$ \\ 
     \hline
     \rule[-7.5pt]{0pt}{20pt}
     $\Psi $ &  CP instrument density & units $1/\mathrm{sr}$ \\
     \hline
     \rule[-7.5pt]{0pt}{20pt}
     $\Psi_s$ &  quantum instrument & none \\
     \hline
     \rule[-7.5pt]{0pt}{20pt}
     $\Psi_{S^2}$ &  quantum channel & none \\ [1ex] 
     \hline
\end{tabular}
    \caption{Linear maps corresponding to Compton scattering.
    Not all have physical meaning as transformations of photon polarization.}
    \label{tab:map_hierarchy}
\end{table}

\textit{POVM density for Compton polarimetry}---
To compute probabilities for Compton scattering experiments, it is convenient to define
\begin{align} \label{eq:POVM_density}
    \Pi(\theta,\phi)
    := \sum_{i=1}^{4} K_i^\dagger K_i = 
    \begin{pmatrix}
        H(\theta) & e^{-2i \phi} G(\theta) \\
        e^{2i\phi} G(\theta) & H(\theta)
    \end{pmatrix} \;.
\end{align}
where $K_i$ are the Kraus operators of $\Psi$ defined in \eqref{eq:Kraus_op_1}-\eqref{eq:Kraus_op_4} (but with the prefactor $\mathcal{N}_0 = \sigma_\mathrm{KN}^{-1/2} \mathcal{N}$ instead of $\mathcal{N}$ because $\Psi$ is normalized), $G(\theta) = \mathcal{N}_0^2 (\Lambda_+^2-\Lambda_-^2) \sin(\xi)$, and $H(\theta) = \mathcal{N}_0^2 (\Lambda_+^2+\Lambda_-^2+r^2)$ with $\mathcal{N}$, $r$, $\xi$, $\Lambda_\pm$ given by \eqref{eq:subKraus1}.
Cyclicity of the trace implies $\mytr(\Psi(\rho)) = \mytr(\Pi(\theta,\phi) \rho)$.
Furthermore, $\int_{S^2} \mathrm{d}\Omega \, \Pi(\theta,\phi) = \id$, so partitioning $S^2$ into $n$ non-overlapping regions $s_n$ gives a positive operator-valued measure (POVM) characterized by the operators $\{\int_{s_i} \mathrm{d}\Omega \, \Pi(\theta,\phi)\}_{i=1}^{n}$.
We thus interpret \eqref{eq:POVM_density} as a \textit{POVM density}.
Note that \eqref{eq:POVM_density} is the same POVM density as derived in Ref.~\cite{Clarke_BellTestPhotons_2026}.

\begin{figure*}
    \includegraphics[width=0.8\textwidth]{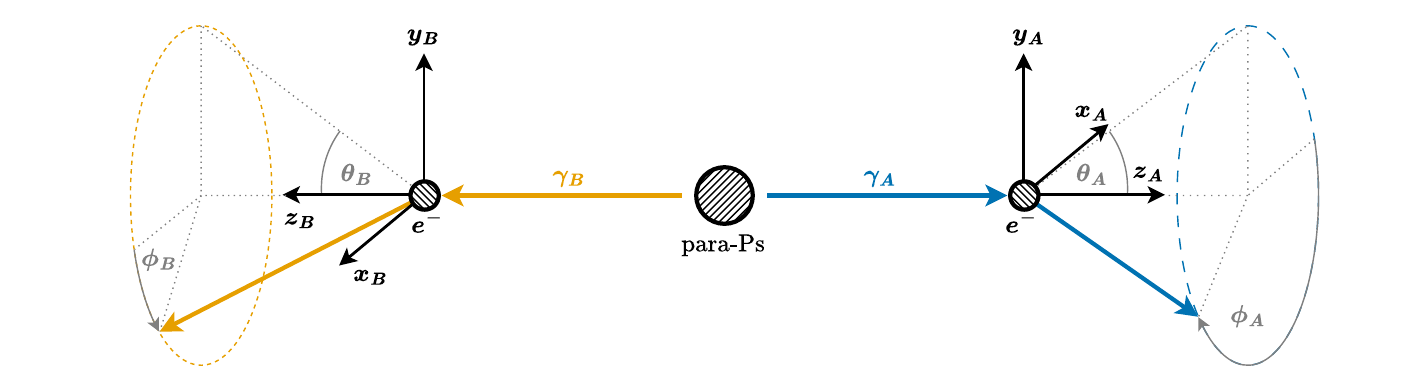}
    \caption{
    Geometry for the decay of para-Positronium at rest into two photons.
    Each photon interacts with a free electron and undergoes Compton scattering.
    The polar and azimuthal scattering angles for photon $\ell$ are $\theta_\ell$ and $\phi_\ell$, respectively, where $\ell\in\{A,B\}$.
    We choose a right-handed coordinate system for each photon based on its direction of propagation.
    }
    \label{fig:para-Ps_decay}
\end{figure*}

\begin{figure}
        \includegraphics[width=0.9\linewidth]{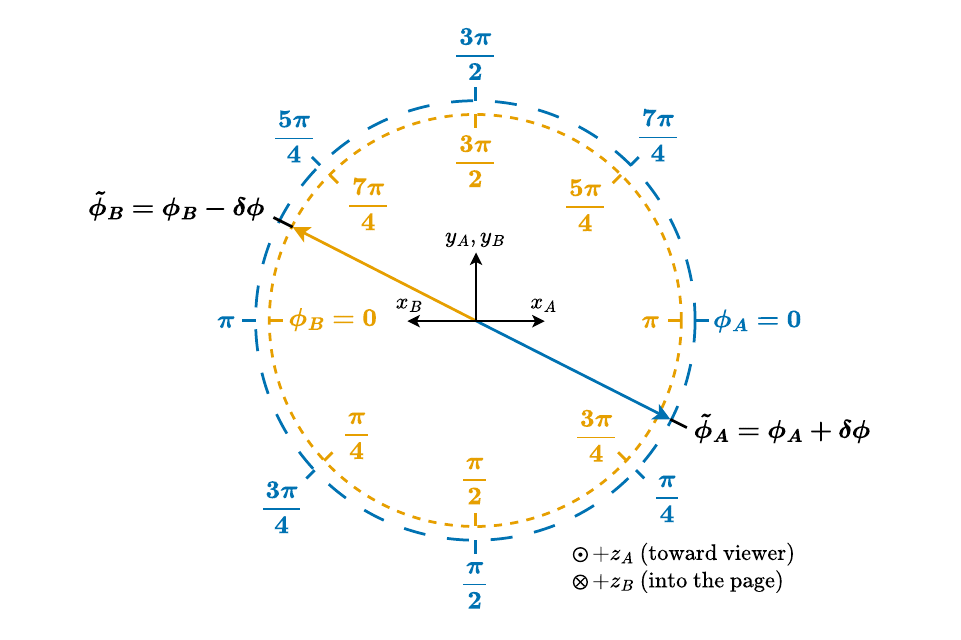}
    \caption{
    Side view of Fig.~\ref{fig:para-Ps_decay} shows our angle conventions for $\phi_A$ (blue) and $\phi_B$ (orange).
    Photon $A$ initially travels along the $z_A$-axis (pointing out of the page) and photon $B$ travels along the $z_B$-axis (pointing into the page).
    A fixed laboratory frame can be used to define the azimuthal offset $\delta\phi = \delta\phi_A = -\delta\phi_B$.
    }
    \label{fig:angle_conventions}
\end{figure}

\textit{Application to para-Ps decay}---
We now consider the decay of para-Ps at rest into two photons, each with energy $E_0 = 1$, and traveling in opposite directions (cf. Fig.~\ref{fig:para-Ps_decay}).
Each photon Compton scatters once, followed by a measurement of its angular distribution.
Crucially, we make no assumptions on the initial polarization of the photon pair and write its joint state as
\begin{align}\label{eq:rhogen}
    \rho = \frac{1}{4} \sum_{i,j=0}^{3} S_{i,j} \, \sigma_i \otimes \sigma_j \;.
\end{align}
Normalization enforces $S_{0,0}=1$.
The reduced states for the individual photons are
\begin{align} \label{eq:reduced_states}
    \rho_A = \mytr_B(\rho) \;, \quad \rho_B = \mytr_A(\rho) \;,
\end{align}
where $\mytr_\ell ( \cdot)$ is the partial trace of subsystem $\ell\in\{A,B\}$.
Let $\theta_\ell$ and $\phi_\ell$ denote the polar and azimuthal scattering angles for photon $\ell$, respectively.
Using \eqref{eq:POVM_density} and \eqref{eq:rhogen}, the probability density per unit solid angle for one Compton scattering on both sides is
\begin{align}
    \frac{\mathrm{d}p_\rho}{\mathrm{d}\Omega_A \mathrm{d}\Omega_B} 
    &= \mytr((\Pi(\theta_A,\phi_A)\otimes \Pi(\theta_B,\phi_B)) \rho) \;.
\end{align}
Alternatively, measuring subsystem $\rho_\ell$ individually is described by
\begin{align}
    \frac{\mathrm{d}p_{\rho_\ell}}{\mathrm{d}\Omega_\ell} = \mytr(\Pi(\theta_\ell,\phi_\ell) \rho_\ell) \;.
\end{align}
In experiments, counting coincidences is proportional to probabilities, i.e.,
\begin{widetext}
\begin{align}\label{eq:q}
    Q(\Tilde{\phi}_A, \Tilde{\phi}_B) &:= \mathcal{F}_{AB}
    \int_{\Tilde{\phi}_A -\Delta\phi_A/2}^{\Tilde{\phi}_A  +\Delta\phi_A/2} \mathrm{d}\phi_A
    \int_{\Tilde{\theta}_A-\Delta\theta_A/2}^{\Tilde{\theta}_A+\Delta\theta_A/2} \mathrm{d}\theta_A
    \int_{\Tilde{\phi}_B-\Delta\phi_B/2}^{\Tilde{\phi}_B+\Delta\phi_B/2} \mathrm{d}\phi_B
    \int_{\Tilde{\theta}_B-\Delta\theta_B/2}^{\Tilde{\theta}_B+\Delta\theta_B/2} \mathrm{d}\theta_B
    \sin\theta_A \sin\theta_B \; \frac{\mathrm{d}p_\rho}{\mathrm{d}\Omega_A \mathrm{d}\Omega_B} \;, \\
     \label{eq:q_i}
     Q_\ell  (\Tilde{\phi}_\ell) 
     &:= \mathcal{F}_\ell
     \int_{\Tilde{\phi}_\ell -\Delta\phi_\ell/2}^{\Tilde{\phi}_\ell  +\Delta\phi_\ell/2} \mathrm{d}\phi_\ell
     \int_{\Tilde{\theta}_\ell-\Delta\theta_\ell/2}^{\Tilde{\theta}_\ell+\Delta\theta_\ell/2} \mathrm{d}\theta_\ell
     \sin\theta_\ell \; \frac{\mathrm{d}p_{\rho_\ell}}{\mathrm{d}\Omega_\ell} \;.
\end{align}
\end{widetext}
The proportionality factors $\mathcal{F}_{AB}$ and $\mathcal{F}_\ell$ are assumed to be angle-independent.
They represent, e.g., detector inefficiencies or source intensity, which might be difficult to determine experimentally.
The angles $\Delta\phi_\ell$ and $\Delta\theta_\ell$ are the azimuthal and polar detector widths for photon $\ell$, respectively.
The central polar angles $\Tilde{\theta}_\ell$ of the detectors are unambiguous relative to the line of flight of the photons, while the central azimuthal angles $\Tilde{\phi}_\ell$ are only defined relative to some arbitrary origin given by the laboratory frame.
In this regard, let $\delta\phi_\ell$ be the difference between $\phi_\ell = 0$ as defined by the polarization basis $\{|H\rangle,|V\rangle\}$ \cite{Supp_Mat}, and $\Tilde{\phi}_\ell =0$ chosen experimentally, i.e., $ \tilde{\phi}_\ell =\phi_\ell + \delta\phi_\ell$.
Even though we might not be able to determine $\delta\phi_A$ and $\delta\phi_B$, we can choose the same origin for both $\phi_A$ and $\phi_B$ so that $\delta\phi = \delta\phi_A = -\delta\phi_B$ (cf. Fig.~\ref{fig:angle_conventions}).

\textit{Entanglement witness for annihilation photons}---
A well-established tool to certify entanglement experimentally are entanglement witnesses, i.e., observables whose expectation values can violate bounds satisfied by all separable states \cite{Chruscinski_Entanglementwitnessesconstruction_2014, Barbieri_DetectionEntanglementPolarized_2003}.
One particular witness for two-qubit systems is the fidelity with any of the four maximally entangled Bell states \cite{Bennett_PurificationNoisyEntanglement_1996}.
However, determining this fidelity for the polarization of a two-photon system generally requires measurements sensitive to circular polarization, which Compton polarimetry is not.
To circumvent this issue, we use a lower bound on the fidelity with one of the Bell states that is independent of circular polarization (see \eqref{eq:lower_bnd_fidelity_phi_minus} in the end matter).
Hence, this bound is in principle accessible in para-Ps experiments using Compton polarimetry.
If the measured result exceeds the critical value of 1/2, the initial photon pair is necessarily entangled in the polarization degree of freedom.
This does not involve reconstructing the full polarization state \footnote{Exact state tomography of the polarization of one photon is impossible using only Compton polarimetry because \eqref{eq:POVM_density} only spans the subspace $\mathrm{span}\{\id, \sigma_x, \sigma_y\}$ and thus any circular polarization component (associated with $\sigma_z$) is inherently unobservable. This also holds for multiple scatterings and/or multiple photons.}.
Furthermore, using this entanglement witness has the advantage that the initial polarization state need not be pure.
Hence, it is robust against moderate decoherence (cf. end matter).

The final difficulty to overcome for an operational proposal is to express this lower bound in \eqref{eq:lower_bnd_fidelity_phi_minus} in terms of quantities that are measurable in para-Ps Compton experiments.
To this end, we define ratios of detector counts for different azimuthal angles in \eqref{eq:ratio1}-\eqref{eq:ratioB} in the end matter, denoted by $\Bar{\mathcal{R}}^{(i)}$ and $\Bar{\mathcal{R}}_\ell^{(i)}$ for $\ell\in\{A, B\}$ and $i\in\{1,2\}$.
The advantage of these ratios over the raw coincidence counts is that the empirical factors $\mathcal{F}_{AB}$ and $\mathcal{F}_\ell$ cancel out.
Using these definitions, we prove our main result in the Supplemental Material \cite{Supp_Mat}: The lower bound on the Bell state fidelity in \eqref{eq:lower_bnd_fidelity_phi_minus} is given by
\begin{align} \label{eq:rho22-rho11_using_ratios}
    \mathcal{I}_{0}
    \left(
    \frac{f(\Bar{\mathcal{R}}^{(1)})-\Bar{\mathcal{R}}_A^{(1)} f(\Bar{\mathcal{R}}^{(2)})}{1+\Bar{\mathcal{R}}_A^{(1)} \Bar{\mathcal{R}}_A^{(2)}} + \frac{1-\Bar{\mathcal{R}}_B^{(1)}}{1+\Bar{\mathcal{R}}_B^{(1)} \Bar{\mathcal{R}}_B^{(2)}}
    \right) \;,
\end{align}
where $f(t) := \frac{t-1}{t+1}$ and $\mathcal{I}_0$ is given in \eqref{eq:I_0} in the end matter.
Hence, if \eqref{eq:rho22-rho11_using_ratios} exceeds 1/2, the initial state is entangled.
We emphasize that \eqref{eq:rho22-rho11_using_ratios} is independent of $\delta\phi$, $S_{1,2}$, and $S_{2,1}$, albeit the individual ratios \eqref{eq:ratio1}-\eqref{eq:ratioB} might not be.
This holds particularly for non-maximally entangled and non-rotationally-invariant states.
Moreover, \eqref{eq:rho22-rho11_using_ratios} does not assume specific detector sizes $\Delta\phi_\ell$ and $\Delta\theta_\ell$.
In fact, these experimental parameters can be chosen freely as long as $\mathcal{I}_0$ remains well-defined.
Thus, the result is adaptable to different experimental architectures, including detectors with low angular resolution.

\textit{Conclusion and Outlook.}---
Our formulation of Compton scattering as a continuous-outcome quantum instrument captures both scattering probabilities and conditional photon polarization states.
It naturally extends to multipartite photonic systems and successive scattering events.
Based on this, we derive a protocol to witness entanglement in para-Ps decay experiments.
In particular, we show how to practically determine a lower bound on the fidelity between the annihilation photons' polarization state and the maximally entangled Bell state they are predicted by QED to be in.
Importantly, the scheme is valid for arbitrary initial two-photon polarization states without assumptions about their structure or purity.

The distinctive feature of our proposal compared to Refs.~\cite{Clarke_BellTestPhotons_2026, Joos_Entangledphotonsparapositronium_2026, Caradonna_Polarizationsymmetrysignature_2026} is its experimental feasibility.
The required correlators can be extracted independently through coarse-grained Compton coincidence measurements using detectors of almost arbitrary angular resolution.
Additionally, the entanglement witness is robust against decoherence effects and independent of factors such as source activity and common detection efficiencies.
This substantially simplifies the experimental requirements.
Consequently, our work establishes a concrete route toward experimentally certifying polarization entanglement in annihilation photons in the near future, without relying on assumptions about the initially emitted quantum state.


\begin{acknowledgments}
\textit{Acknowledgments.}---
This research was funded in whole or in part by the Austrian Science Fund (FWF) [10.55776/COE1].
This research was supported by the Austrian Ministry for Women, Science and Research (project WTZ PL 11/2026).
\end{acknowledgments}


\bibliography{references}

\section*{End Matter}

\subsection{Fano and Müller matrices}
The Fano scattering matrix $T$ and the Müller rotation matrix $M$ are given by \cite{Fano_RemarksClassicalQuantumMechanical_1949, McMaster_MatrixRepresentationPolarization_1961}
\begin{align} \label{eq:T-matrix}
    T &= t_{00}
    \begin{pmatrix}
        t_{11} & t_{12} & 0 & 0 \\
        t_{12} & 2-t_{12} & 0 & 0 \\
        0 & 0 & t_{33} & 0 \\
        0 & 0 & 0 & t_{44}
    \end{pmatrix} \;, \\ 
    M &= \begin{pmatrix} \label{eq:M-matrix}
        1 & 0 & 0 & 0 \\
        0 & \cos2\phi & \sin2\phi & 0 \\
        0 & -\sin2\phi & \cos2\phi & 0 \\
        0 & 0 & 0 & 1 
    \end{pmatrix} \;.
\end{align}
Here,
\begin{align}
    t_{00} &= \frac{r_e^2}{2} \; \left(\frac{E_1}{E_0}\right)^2 \;,\label{eq:t00} \quad
    t_{12} = \sin^2\theta \;, \quad
    t_{33} = 2\cos\theta \;,\\
    t_{11} &= \frac{E_0}{E_1} + \frac{E_1}{E_0} - \sin^2\theta \;, \quad
    t_{44} = (\frac{E_0}{E_1}+\frac{E_1}{E_0}) \cos\theta \;,
    \label{eq:t44}
\end{align}
where $r_e$ is the classical electron radius, and the post-scattering photon energy
\begin{align}\label{eq:energy_E1}
    E_1 =\frac{E_0}{1+E_0(1-\cos\theta)} \;.
\end{align}
Furthermore, $\theta$ and $\phi$ are the polar and azimuthal scattering angles, respectively.
The angle $\phi$ increases clockwise, i.e., from $|V\rangle$ to $|H\rangle$ (cf. Fig.~\ref{fig:angle_conventions} and the Supplementary Material \cite{Supp_Mat}).

\subsection{Natural representation of $\Phi$}
Consider the matrix $X = \sum_{i,j=0}^{1} x_{i,j} |i\rangle\langle j| = \frac{1}{2} \sum_{k=0}^{3} S_k^{(x)} \sigma_k$
with $S_k^{(x)} = \mytr(X \sigma_k)$.
Using $|S_x\rangle = \sum_{k=0}^{3} S_k^{(x)} |k\rangle$ and $|X\rrangle = \sum_{i,j=0}^{1} x_{i,j} |i\rangle\otimes |j\rangle$, it can be checked that $|S_x\rangle = C|X\rrangle$ holds with
\begin{align} \label{eq:C-matrix}
    C = \begin{pmatrix}
        1 & 0 & 0 & 1 \\
        0 & 1 & 1 & 0 \\
        0 & i & -i & 0 \\
        1 & 0 & 0 & -1 \\
    \end{pmatrix} \;.
\end{align}
Consequently, we can derive the natural representation of $\Phi$ as described in the main text, giving
\begin{align} \label{eq:nat_rep_of_Phi}
    F(\Phi) = C^{-1} TM C
 = \kappa 
    \begin{pmatrix}
    \alpha & e^{2i\phi} xy & e^{-2i\phi} xy & \beta \\
    xy & e^{2i\phi}x^2 & e^{-2i\phi} y^2 & xy \\
    xy & e^{2i\phi} y^2 & e^{-2i\phi} x^2 & xy \\
    \beta & e^{2i\phi} xy & e^{-2i\phi} xy & \alpha \\
    \end{pmatrix} \;,
\end{align}
where $x = 1+\cos\theta$, $y = 1-\cos\theta$, $\kappa = t_{00} / 2$, $\alpha = t_{11}+t_{44}$, and $\beta  = t_{11}-t_{44}$ with $t_{i,j}$ in \eqref{eq:t00}-\eqref{eq:t44}.

\subsection{Choi representation of $\Phi$ and complete positivity}
Using the realignment operator $\mathcal{R}$, acting on the computational basis states as $\mathcal{R}(|i\rangle\langle j| \otimes |k\rangle\langle l|) = |i\rangle\langle k| \otimes |j\rangle\langle l|$,
the Choi representation of $\Phi$ is $J(\Phi) = \mathcal{R}(F(\Phi))$ \cite{Watrous_TheoryQuantumInformation_2018}.
The map $\Phi$ is completely positive if and only if $J(\Phi)$ is positive semidefinite.
Hence, unitarily changing the basis does not alter this property.
Considering $U=\mathrm{diag}(1,e^{2i\phi}, 1, e^{2i\phi})$, we find
\begin{align} \label{eq:CPtest_matrix}
    U J(\Phi) U^\dagger = 
    \kappa
    \begin{pmatrix}
     \alpha   & xy   & xy   &  x^2 \\
xy    & \beta   &  y^2  & xy   \\
xy   &  y^2  & \beta   & xy  \\
x^2  & xy   & xy   & \alpha 
    \end{pmatrix} \;.
\end{align}
Thus, complete positivity of $\Phi$ is independent of $\phi$.
A direct computation shows that two of the eigenvectors of \eqref{eq:CPtest_matrix} are the Bell states $\frac{1}{\sqrt{2}}(1,0,0,-1)^T$ and $\frac{1}{\sqrt{2}}(0,1,-1,0)^T$, with respective eigenvalues
\begin{align}
    \alpha-x^2 &= \frac{E_0^2 (1-\cos\theta)^2(1+\cos\theta)}{1+E_0(1-\cos\theta)} \;,\\
    \beta-y^2 &= \frac{E_0^2 (1-\cos\theta)^3}{1+E_0(1-\cos\theta)} \;.
\end{align}
Both are nonnegative for the physically meaningful parameter region $E_0>0$ and $0\leq \theta\leq \pi$.
Consequently, to investigate whether $J(\Phi) \geq 0$, we can project \eqref{eq:CPtest_matrix} onto the orthogonal complement of the subspace spanned by the two eigenvectors above, giving
\begin{align}
    \begin{pmatrix}
        \alpha + x^2 & 2 x y \\
        2 x y & \beta + y^2
    \end{pmatrix}
     = 
     2 \begin{pmatrix} x \\ y\end{pmatrix} \begin{pmatrix} x & y\end{pmatrix} + \frac{E_0^2 y^2}{1+E_0 y} \begin{pmatrix} x & 0 \\ 0 & y\end{pmatrix} \;.
\end{align}
For $E_0 >0$ and $0\leq \theta \leq \pi$, we have $x\geq0$ and $y\geq0$, so both sides of the equation are positive semidefinite.
Hence, $J(\Phi)$ is positive semidefinite and $\Phi$ is completely positive.

\subsection{Kraus representation of $\Phi$}
Computing the eigenvectors of $J(\Phi)$ yields the Kraus operators \cite{Watrous_TheoryQuantumInformation_2018}
\begin{align} \label{eq:Kraus_op_1}
    K_1 &= \mathcal{N}
    \begin{pmatrix}
        -r \cos(\theta/2) & 0\\
        0 & r e^{-2i\phi} \cos(\theta/2)\\
    \end{pmatrix} \;,\\
    K_2 &= \mathcal{N}
    \begin{pmatrix}
        0 & -r e^{-2i\phi} \sin(\theta/2) \\
        r \sin(\theta/2) & 0 \\
    \end{pmatrix} \;,\\
    K_3 &= \mathcal{N} \Lambda_-
    \begin{pmatrix}
         \sin(\xi/2) & -e^{-2i\phi} \cos(\xi/2) \\
         -\cos(\xi/2) & e^{-2i\phi} \sin(\xi/2)\\
    \end{pmatrix} \;,\\
    K_4 &= \mathcal{N} \Lambda_+
    \begin{pmatrix}
        \cos(\xi/2) & e^{-2i\phi} \sin(\xi/2) \\
        \sin(\xi/2) & e^{-2i\phi} \cos(\xi/2)\\
    \end{pmatrix} \;,\label{eq:Kraus_op_4}
\end{align}
with
\begin{equation}
\begin{aligned}\label{eq:subKraus1}
    \mathcal{N} &= 2\sqrt{\kappa} = r_e \frac{E_1}{E_0} \;, \quad
    r = \frac{(1-\cos\theta)\sqrt{E_0 E_1}}{2} \;,\\
    \xi &= \mathrm{arctan2}\left( 2(1+r^2)\cos\theta \;, \; \sin^2\theta \right) \;,\\
    \Lambda_\pm &= \frac{1}{2} \sqrt{1+\cos^2\theta+2r^2 \pm \sqrt{4(1+r^2)^2\cos^2\theta+\sin^4\theta}} \;,
\end{aligned}
\end{equation}
where $E_1$ is given in \eqref{eq:energy_E1}, and $\mathrm{arctan2}(t,u)$ is the arctangent of $u/t$, taking into account which quadrant the point $(t,u)$ is in.
The map $\Phi$ acts as
\begin{align}
    \Phi(\rho) = \sum_{i=1}^{4} K_i \, \rho \, K_i^\dagger \;.
\end{align}
The role of $\phi$ for $\Phi$ is elucidated by defining $U_\phi := \mathrm{diag}(1,e^{-2i \phi})$, with which we can write
\begin{align} \label{eq:Phi_Kraus_TildeK_and_Uphi}
    \Phi(\rho) = \sum_{i=1}^{4} \Tilde{K}_i(\theta) \, U_\phi \, \rho \, U_\phi^\dagger \, \Tilde{K}_i^\dagger(\theta) \;,
\end{align}
where $\Tilde{K}_i(\theta) = K_i U_\phi^\dagger$.
Thus, $\Phi$ acts as a unitary rotation depending only on $\phi$, followed by the linear map defined by the Kraus operators $\Tilde{K}_i(\theta)$, which depend only on $\theta$.
This mirrors the Stokes vector transformation \eqref{eq:stokes_trafo}, where $|S\rangle$ is transformed by first applying the Müller rotation matrix $M$ followed by the Fano transformation matrix $T$.

\subsection{Bounding the fidelity with the Bell state $|\Phi^-\rangle$}
We can lower bound the fidelity of the general bipartite qubit state in \eqref{eq:rhogen} with the maximally entangled state $|\Phi^-\rangle$ as follows.
With $|\Phi^\pm\rangle = \frac{1}{\sqrt{2}}(|R\rangle\otimes |R\rangle \pm |L\rangle\otimes |L\rangle)$ we obtain
\begin{align}\label{eq:F_Phi^-}
    F_{\Phi^-} &= \langle\Phi^-|\varrho|\Phi^-\rangle = \frac14\left( 1-S_{1,1}+S_{2,2}+S_{3,3} \right) \;. \\
    F_{\Phi^+} &= \langle\Phi^+|\varrho|\Phi^+\rangle = \frac14\left( 1+S_{1,1}-S_{2,2}+S_{3,3} \right) \;. \label{eq:F_Phi^+}
\end{align}
If either exceeds 1/2, the system is entangled \cite{Bennett_PurificationNoisyEntanglement_1996}.
Furthermore,
\begin{align} \label{eq:fidelity_intermediate_eq}
    F_{\Phi^-} -\frac{S_{2,2}-S_{1,1}}{2} = F_{\Phi^+} \;.
\end{align}
Since $\rho\geq 0$ implies $F_{\Phi^+} \geq 0$, we get
\begin{align} \label{eq:lower_bnd_fidelity_phi_minus}
    F_{\Phi^-} \geq \frac{S_{2,2}-S_{1,1}}{2} \;.
\end{align}
This lower bound of $F_{\Phi^-}$ is an entanglement witness: If it is larger than 1/2, the state is necessarily entangled.
Crucially, it is independent of the circular polarization part of $\rho$, corresponding to the matrix elements $S_{i,j}$ with $i=3$ or $j=3$ in \eqref{eq:rhogen}.

QED predicts the polarization state of annihilation photons to be $|\Phi^-\rangle$.
Here, we consider the isotropic state $\rho_\mathrm{iso} = \frac{1-p}{4} \id + p |\Phi^-\rangle\langle \Phi^-|$ with $-\frac{1}{3} \leq p \leq 1$, which is entangled for $1/3<p\leq 1$ \cite{Horodecki_Reductioncriterionseparability_1999}.
Its fidelity with $|\Phi^-\rangle$ is $F_{\Phi^-} = \frac{1+3p}{4}$, which reaches $F_{\Phi^-} = 1/2$ at exactly $p=1/3$.
Writing $\rho_\mathrm{iso}$ in the form of \eqref{eq:rhogen} reveals $S_{0,0} = 1$ and $S_{1,1} =- S_{2,2} = -S_{3,3}= -p$, and \eqref{eq:lower_bnd_fidelity_phi_minus} implies $F_{\Phi^-} \geq p$.
Consequently, even if decoherence in annihilation photons degrades the maximal entanglement to the mixed state $\rho_\mathrm{iso}$, the bound \eqref{eq:lower_bnd_fidelity_phi_minus} still certifies entanglement for $1/2<p\leq 1$.

\subsection{Definitions for measuring the entanglement witness}
For witnessing entanglement with \eqref{eq:rho22-rho11_using_ratios} experimentally, one needs to measure the ratios
\begin{align}
    \Bar{\mathcal{R}}^{(1)} &= \frac{Q(\delta\phi, \pi/2-\delta\phi)}{Q(\delta\phi, -\delta\phi)}\;,\label{eq:ratio1}\\
    \Bar{\mathcal{R}}^{(2)} &= \frac{Q(\pi/4+\delta\phi, 3\pi/4-\delta\phi)}{Q(\pi/4+\delta\phi, \pi/4-\delta\phi)} \;,\label{eq:ratio2}\\
    \Bar{\mathcal{R}}_A^{(1)} &= \frac{Q_A(\pi/4+\delta\phi)}{Q_A(\delta\phi)} \;, \;
    \Bar{\mathcal{R}}_A^{(2)} = \frac{Q_A(\pi/2+\delta\phi)}{Q_A(\pi/4+\delta\phi)} \;,\label{eq:ratioA}\\
    \Bar{\mathcal{R}}_B^{(1)} &= \frac{Q_B(\pi/4-\delta\phi)}{Q_B(-\delta\phi)}\;, \;
    \Bar{\mathcal{R}}_B^{(2)} = \frac{Q_B(\pi/2-\delta\phi)}{Q_B(\pi/4-\delta\phi)} \;.\label{eq:ratioB}
\end{align}
For the prefactor in \eqref{eq:rho22-rho11_using_ratios}, we have
\begin{align}
    \mathcal{I}_0 &= \frac{\Delta\phi_A \Delta\phi_B}{\sin\Delta\phi_A\sin\Delta\phi_B}\frac{\mathcal{I}_{H_A} \mathcal{I}_{H_B}}{\mathcal{I}_{G_A} \mathcal{I}_{G_B}} \;, \label{eq:I_0}\\
    \mathcal{I}_{G_\ell} &= \int_{\Tilde{\theta}_\ell-\Delta\theta_\ell/2}^{\Tilde{\theta}_\ell+\Delta\theta_\ell/2} \sin\theta_\ell \mathrm{d}\theta_\ell G(\theta_\ell) \;, \label{eq:I_G_i}\\
    \mathcal{I}_{H_\ell} &= \int_{\Tilde{\theta}_\ell-\Delta\theta_\ell/2}^{\Tilde{\theta}_\ell+\Delta\theta_\ell/2} \sin\theta_\ell \mathrm{d}\theta_\ell H(\theta_\ell) \;, \label{eq:I_H_i}
\end{align}
which are fully determined by the experimental parameters $\Tilde{\theta}_\ell$, $\Delta\theta_\ell$, $\Delta\phi_\ell$ and $E_0$.
Furthermore, $\mathcal{I}_{G_\ell} \neq 0$ for $\tilde{\theta}_\ell \neq n \pi$ and $\Delta\theta_\ell\neq 2n\pi$ with $n\in\mathbbm{Z}$, and $\sin\Delta\phi_\ell\neq 0$ for $\Delta\phi_\ell\neq n\pi$ with $n\in\mathbbm{Z}$.
Thus, $\mathcal{I}_0$ is well-defined for almost all detector sizes.


\clearpage
\onecolumngrid
\setcounter{equation}{0}
\setcounter{figure}{0}
\setcounter{table}{0}
\setcounter{section}{0}
\setcounter{page}{1}
\renewcommand{\theequation}{S\arabic{equation}}
\renewcommand{\thefigure}{S\arabic{figure}}
\renewcommand{\thetable}{S\arabic{table}}
\renewcommand{\thesection}{\arabic{section}}
\renewcommand{\theHequation}{S\arabic{equation}}
\renewcommand{\theHfigure}{S\arabic{figure}}
\renewcommand{\theHtable}{S\arabic{table}}
\renewcommand{\theHsection}{S\arabic{section}}
\setcounter{secnumdepth}{2}

\begin{center}
    {\large \textbf{Supplemental Material: \\[1mm]
   Witnessing Polarization Entanglement of Annihilation Photons via Compton Scattering}}
\end{center}

\begin{centering}
Tobias C. Sutter\,\orcidlink{0009-0008-3179-7979},$^{1}$~
Beatrix C. Hiesmayr\,\orcidlink{0000-0001-9062-6039},$^{1,2}$\\
\end{centering}

\vspace{2mm}

\begin{centering}
\textit{\small
$^1$ University of Vienna, Faculty of Physics, Währingerstrasse 17, 1090 Vienna, Austria. \\
$^2$ IT:U Interdisciplinary Transformation University, Freistädter Strasse 400, 4040 Linz, Austria. \\
}
\end{centering}

%
\begin{quote}
{\small{
In this Supplementary Material, we first discuss why previously proposed entanglement witnesses are currently experimentally unfeasible.
Then, we briefly explain our angle conventions for para-Ps decay, followed by a basic review of linear map representations in quantum information theory.
Using the formulation of Compton scattering as a quantum instrument density derived in the main text, we discuss how to treat multiple successive scatterings.
Finally, we derive how to determine the lower bound of the fidelity with a Bell state using only experimentally accessible ratios of probabilities, and give specific numerical values for a proposed experimental realization.
}}
\end{quote}


\section{Experimental feasibility of entanglement witnesses for annihilation photons from the literature}

Other theoretical investigations have identified alternative signatures of polarization entanglement in annihilation photons.
Recently, the authors of Ref.~\cite{Clarke_BellTestPhotons_2026} proposed a way to perform a Bell test for annihilation photons using Compton polarimetry.
In particular, with their setup, they claim to be able to exclude any local hidden-variable description when each photon undergoes at least two Compton scatterings.
Note that this type of nonlocality is more restrictive than entanglement: Every nonlocal state is entangled, but not every entangled state is nonlocal \cite{Werner_QuantumstatesEinsteinPodolskyRosen_1989}.
However, the experimental crux is that multiple scatterings at very specific angles are required.
While the authors say a more detailed feasibility study is ongoing, they present preliminary results using their own Geant4 simulation.
They predict that only about 300 coincidence events per week are expected at a source activity of 1 GBq, and that even under optimal conditions, measurement times of several months would likely be required.
More importantly, however, they do not establish whether a real readout system can reliably identify the relevant events under the proposed high-flux operating conditions.

Ref.~\cite{Caradonna_Polarizationsymmetrysignature_2026} establishes an equivalence between polarization symmetry and maximally entangled Bell states using single-Compton scattering.
A limitation of their scheme is its restriction to pure, maximally entangled states.
Since environmental decoherence can render the polarization state mixed without necessarily destroying its entanglement, exact polarization symmetry may be lost even for genuinely entangled photon pairs.
Consequently, the criterion does not provide a robust entanglement witness for realistic experimental conditions, where imperfect state preparation and decoherence must be accounted for.
The authors acknowledge the experimental challenges and that an explicit experimental reconstruction protocol remains to be developed.

Lastly, Ref.~\cite{Joos_Entangledphotonsparapositronium_2026} proposes a measurement scheme based on coincidence correlations in different polarization bases.
In particular, they compute ratios similar to \eqref{eq:ratio1}-\eqref{eq:ratioB} but with infinite angular precision and show that specific correlations uniquely identify the initial Bell state.
They also discuss extending their results to finite angle intervals, but note that entanglement certification becomes harder as detector sizes increase, because the signatures of entangled and separable states become increasingly similar.


\section{Note on the polarization basis and angle convention}

Consider the Stokes vector $|S\rangle = \sum_{i=0}^{3} S_i |i\rangle$ of a photon with polarization $\rho$, defined by $S_i = \langle\sigma_i\rangle_\rho = \mytr(\rho \sigma_i)$, where $\sigma_0 = \id$, and $\sigma_i$ for $i\in\{1,2,3\}$ are the Pauli matrices.
We assign (cf. Ref.~\cite{McMaster_MatrixRepresentationPolarization_1961} Table I; Fig.~\ref{fig:angle_polarization_conventions} below)
\begin{align}
    S_1 = +1 \Leftrightarrow |V\rangle \;,\; S_1 = -1 \Leftrightarrow |H\rangle \;, \;
    S_2 = +1 \Leftrightarrow |-45^\circ\rangle  \;,\; S_2 = -1 \Leftrightarrow |+45^\circ\rangle \;,\;
    S_3 = +1 \Leftrightarrow |L\rangle \;,\; S_3 = -1 \Leftrightarrow |R\rangle \;.
\end{align}
In particular, with the computational basis vectors defined as $|0\rangle = |L\rangle = (1,0)^T$ and $|1\rangle = |R\rangle = (0,1)^T$, the other basis states are
\begin{align}
    |V\rangle = \frac{1}{\sqrt{2}} \begin{pmatrix} 1 \\ 1 \end{pmatrix} \;, \quad
    |H\rangle = \frac{1}{\sqrt{2}} \begin{pmatrix} -i\\ i \end{pmatrix} \;,\quad
    |-45^\circ\rangle = \frac{|H\rangle+|V\rangle}{\sqrt{2}} \;,\quad 
    |+45^\circ\rangle = \frac{|H\rangle - |V\rangle}{\sqrt{2}} \;.
\end{align}
Note that the global phase in $|H\rangle$ needs to be retained so that the equations for $|-45^\circ\rangle$ and $|+45^\circ\rangle$ are valid.

\begin{figure}
    \centering
    \includegraphics[width=0.35\linewidth]{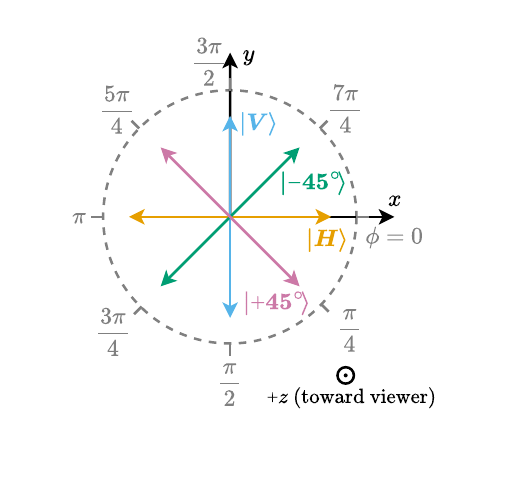}
    \caption{
    Angle conventions corresponding to the polarization basis states.
    Left and right circular polarization correspond to anticlockwise and clockwise rotations, respectively.
    }
    \label{fig:angle_polarization_conventions}
\end{figure}

Besides keeping the specific forms of the Fano scattering matrix $T$ and the Müller rotation matrix $M$ from Ref.~\cite{McMaster_MatrixRepresentationPolarization_1961} valid, a different argument for this convention can be made by considering the Thomson limit ($E_0 \rightarrow 0$) of the Klein-Nishina cross section density per unit solid angle,
\begin{align}
    \frac{\mathrm{d}\sigma_T}{\mathrm{d}\Omega} &= \lim_{E_0\rightarrow0} S_0' = \lim_{E_0\rightarrow0} \frac{\mathrm{d}\sigma_{\rho}}{\mathrm{d}\Omega} =  \lim_{E_0\rightarrow0} t_{00} (t_{11} S_0 + t_{12} (S_1 \cos2\phi + S_2 \sin2\phi)) \\
    &= \frac{r_e^2}{2} \left( (1+\cos^2\theta) S_0 + \sin^2\theta( S_1 \cos(2\phi) + S_2 \sin(2\phi)) \right) \;.
\end{align}
Evaluating this for a normalized quantum state ($S_0=1$) and at $\theta=\pi/2$, $\phi=0$ yields
\begin{align} \label{eq:thomson1}
    \frac{\mathrm{d}\sigma_T}{\mathrm{d}\Omega} \vert_{\theta=\pi/2, \phi=0} =\frac{r_e^2}{2} (1+S_1) \;.
\end{align}
On the other hand, a classical calculation gives \cite{Heitler_quantumtheoryofradiation}
\begin{align}
    \frac{\mathrm{d}\sigma_T}{\mathrm{d}\Omega} = r_e^2 \sin^2\chi \;,
\end{align}
where $\chi$ is the angle between the scattering direction $\Vec{R}$ and the direction of polarization of the incident photon $\Vec{E}$.
We take horizontal and vertical polarization $H$ and $V$ along the $x$ and $y$ axes, respectively, and the incident photon direction along $z$ (cf. Fig.~\ref{fig:angle_polarization_conventions}).
Then, for $H$-polarized radiation, we have $\Vec{E} = (1,0,0)$.
The outgoing photon direction can be parametrized by $\Vec{R} = (\sin\theta \cos\phi,- \sin\theta \sin\phi, \cos\theta)$, consistent with the azimuthal angle $\phi$ increasing clockwise.
Hence, $\cos\chi = \Vec{E}\cdot\Vec{R} = \sin\theta\cos\phi$, and thus
\begin{align}
    \frac{\mathrm{d}\sigma_T}{\mathrm{d}\Omega} = r_e^2 (1-\sin^2\theta \cos^2\phi) \;.
\end{align}
For the scattering angles $\theta=\pi/2$ and $\phi=0$, this becomes
\begin{align} \label{eq:thomson2}
    \frac{\mathrm{d}\sigma_T}{\mathrm{d}\Omega} \vert_{\theta=\pi/2, \phi=0} = 0 \;.
\end{align}
Comparing \eqref{eq:thomson1} and \eqref{eq:thomson2}, we find that horizontally polarized photons correspond to $S_1 = \mytr(\rho\sigma_x) = -1$.
Similar considerations recover $S_1 = +1$ for $|V\rangle$ with $\phi=\pi/2$, $S_2 = -1$ for $|+45^\circ\rangle$ with $\phi = \pi/4$, and $S_2 = +1$ for $|-45^\circ\rangle$ with $\phi = -\pi/4$.


\section{Linear quantum maps and their representations}
\label{app:linear_maps_representations}

The main reference for this section is Ref.~\cite{Watrous_TheoryQuantumInformation_2018}.
Let $\hs_1, \hs_2$ be finite-dimensional Hilbert spaces, and $\linop(\hs_1,\hs_2)$ the set of linear operators mapping $\hs_1$ into $\hs_2$.
The special case $\hs_1=\hs_2$ is simply denoted by $\linop(\hs_1)$.
The set of density operators on $\hs_1$ is defined as $\denop(\hs_1) := \{ A \in \linop(\hs_1) \,| \, A \geq 0, \mytr(A) = 1 \}$.

\subsection{Natural representation}
Define $d_1=\dim(\hs_1)$ and $d_2=\dim(\hs_2)$, and let $\{|i\rangle_1\}_{i=0}^{d_1-1} \subset \hs_1$ and $\{|i\rangle_2\}_{i=0}^{d_2-1} \subset \hs_2$ be the computational bases of $\hs_1$ and $\hs_2$, respectively.
Then we can expand $\rho=\sum_{i,j=0}^{d_1-1} \rho_{i,j} |i\rangle_1\langle j|_1$ and $\Phi(\rho) = \sum_{i,j=0}^{d_2-1} \Phi(\rho)_{i,j} |i\rangle_2\langle j|_2$.
Define the vectorization map $\mathrm{vec}:\linop(\hs_\ell) \mapsto \hs_\ell\otimes\hs_\ell$ for $\ell\in\{1,2\}$ by its action on the computational basis,
\begin{align}
    \mathrm{vec}(|i\rangle_\ell\langle j|_\ell) = |i\rangle_\ell\otimes |j\rangle_\ell \;.
\end{align}
This is a linear bijection, i.e., every vector $|A\rangle\in\hs_\ell\otimes\hs_\ell$ uniquely determines an operator $A\in\linop(\hs_\ell)$.
We denote $\mathrm{vec}(A) \equiv |A \rrangle$.
For every linear map $\Phi:\linop(\hs_1)\rightarrow\linop(\hs_2)$, there exists a linear operator $F(\Phi):\linop(\hs_1\otimes\hs_1, \hs_2\otimes\hs_2)$ such that
\begin{align}
    F(\Phi) |A\rrangle = |\Phi(A)\rrangle
\end{align}
for all $A\in\linop(\hs_1)$.
The operator $F(\Phi)$ is called the \textit{natural representation of $\Phi$}.

\subsection{Choi representation}

Define the realignment operator $\mathcal{R}:\linop(\hs_1\otimes\hs_1, \hs_2\otimes \hs_2) \rightarrow \linop(\hs_2\otimes \hs_1)$ by its action on the computational basis states,
\begin{align} \label{eq:realignment_operator_supp_mat}
    \mathcal{R}(|i\rangle_2\langle j|_1 \otimes |k\rangle_2\langle l|_1) = |i\rangle_2\langle k|_2 \otimes |j\rangle_1\langle l|_1 \;.
\end{align}
It is straightforward to show $\mathcal{R}(A\otimes B^\ast) = |A\rrangle\llangle B|$, where $A,B\in\linop(\hs_1,\hs_2)$ and $^*$ denotes complex conjugation.
Applying $\mathcal{R}$ to $F(\Phi)$ yields the \textit{Choi representation of $\Phi$},
\begin{align}
    J(\Phi) = \mathcal{R}(F(\Phi)) = \sum_{i,j=0}^{d_1-1} \Phi(|i\rangle_1\langle j|_1) \otimes |i\rangle_1\langle j|_1 \;.
\end{align}
Both the natural and Choi representations determine the linear map $\Phi$ \textit{uniquely}.

\subsection{Kraus representation}

Using the singular value decomposition, we can write
\begin{align}
    J(\Phi) = \sum_{i=1}^{r_K} |K_i\rrangle\llangle L_i| \;,
\end{align}
where $1\leq r_k = \mathrm{rank}(J(\Phi)) \leq d_1 d_2$.
The number $r_K$ is known as the Kraus rank of $\Phi$.
It holds that
\begin{align} \label{eq:kraus_def}
    \Phi(\rho) = \sum_{i=1}^{r_K} K_i A L_i^\dagger
\end{align}
for all $A\in\linop(\hs_1)$.
The collection $\{K_i\}_{i=1}^{r_K}$ together with $\{L_i\}_{i=1}^{r_K}$ is known as the \textit{Kraus representation of $\Phi$}.
If $\Phi$ preserves hermiticity, then $J(\Phi)$ is hermitian and thus $|K_i\rrangle=|L_i\rrangle$, so the Kraus representation is given by $\{K_i\}_{i=1}^{r_K}$.

The Kraus representation is not unique.
To see this, consider the unitary matrix $U=\sum_{i,j=1}^{r_K} u_{i,j} |i\rangle\langle j|$.
It holds that $\sum_{i=1}^{r_K} u_{i,j} u_{i,l}^\ast = \delta_{j,l}$.
Then, the two sets of operators $\{K_i\}_{i=1}^{r_K}$ and $\{K'_i = \sum_{j=1}^{r_K} u_{i,j} K_j\}_{i=1}^{r_K}$ characterize the same linear map because
\begin{align}
    \sum_{i=1}^{r_K} K'_i A (K'_i)^\dagger = \sum_{i,j,l=1}^{r_K} u_{i,j} u_{i,l}^\ast K_j A K_l^\dagger = \sum_{i=1}^{r_K} K_i A K_i^\dagger
\end{align}
for all $A\in \linop(\hs_1)$.

\subsection{Properties of linear maps}

Let the set of positive semidefinite operators on $\hs$ be denoted by $\mathrm{Pos}(\hs) = \{ AA^\dagger : A \in\linop(\hs)\}$, where $A^\dagger$ is the adjoint of $A$.
We say a linear map $\Phi:\linop(\hs_1)\rightarrow\linop(\hs_2)$ is \textit{positive} if $A\in\mathrm{Pos}(\hs_1)$ implies $\Phi(A) \in\mathrm{Pos}(\hs_2)$.
Additionally, $\Phi$ is \textit{completely positive} if the map $\Phi\otimes \id_{\linop(\hs)}$ is positive for every Hilbert space $\hs$, where $\id_{\linop(\hs)}$ denotes the identity map on $\linop(\hs)$.
Importantly, $\Phi$ is completely positive if and only if $J(\Phi)\in\mathrm{Pos}(\hs_2\otimes \hs_1)$, wher $J(\Phi)$ is its Choi representation.

The linear map $\Phi:\linop(\hs_1)\rightarrow\linop(\hs_2)$ is called \textit{trace non-increasing} if $\mytr(\Phi(A)) \leq \mytr(A)$ for all $A\in\mathrm{Pos}(\hs_1)$.
If equality is attained, the map is called \textit{trace-preserving}.
Let $\{K_i\}_{i=1}^{r_K}$ and $\{L_i\}_{i=1}^{r_K}$ be the Kraus operators of $\Phi$.
The map $\Phi$ is trace non-increasing if and only if
\begin{align} \label{eq:TP_kraus}
    \sum_{i=1}^{r_K} L_i^\dagger K_i \leq \id_{d_1} \;,
\end{align}
where the inequality is understood in the positive semidefinite ordering, i.e., it is satisfied if and only if $\id_{d_1}-\sum_{i=1}^{r_K} L_i^\dagger K_i \in \mathrm{Pos}(\hs_1)$.
For trace preservation, equality holds in \eqref{eq:TP_kraus}.


\section{Successive Compton scatterings}
\label{app:multiple_scatterings}
We consider a photon with energy $E_0$ which undergoes successive Compton scatterings.
As discussed in the main text, one scattering on an unpolarized electron at rest, with the final electron spin unobserved, can be described by the normalized instrument density $\Psi = \sigma_\mathrm{KN}^{-1} \Phi$ as in \eqref{eq:psi_instrument_density}.
The total cross section $\sigma_\mathrm{KN}$ in \eqref{eq:sigma_tot_2} can be computed to be
\begin{align}
    \sigma_\mathrm{KN} = \frac{\pi \, r_e^2}{E_0^3} \left( \frac{2E_0 (2+8E_0+9E_0^2+E_0^3)}{(1+2E_0)^2} + (E_0^2-2(E_0+1)) \log(1+2E_0) \right) \;.
\end{align}
We observe that $\sigma_\mathrm{KN} = \sigma_\mathrm{KN}(E_0)$, i.e., it depends on the incoming photon energy $E_0$ (and the classical electron radius $r_e$).
To make this explicit, we write the instrument density as
\begin{align}
    \Psi_{E_0,\Omega_0} = \frac{1}{\sigma_\mathrm{KN}(E_0)} \Phi_{E_0,\Omega_0} \;,
\end{align}
where we also made explicit the dependence of the map on the scattering angle $\Omega_0 = (\theta_0, \phi_0)$.
We have
\begin{align}
    \int_{S^2} \mathrm{d}\Omega_0 \; \mytr(\Psi_{E_0,\Omega_0}(\rho)) = \mytr(\rho) \;.
\end{align}
After the first scattering, the photon's energy changes as
\begin{align}
    E_0 \rightarrow E_1(E_0, \Omega_0) = \frac{E_0}{1+E_0(1-\cos\theta_0)} \;.
\end{align}
Therefore, a subsequent second scattering is described by
\begin{align}
    \Psi_{E_1, \Omega_1} = \frac{1}{\sigma_\mathrm{KN}(E_1(E_0, \Omega_0))} \Phi_{E_1,\Omega_1} \;.
\end{align}
Concatenating the instrument densities for two successive scatterings gives the probability density per unit solid angle
\begin{align}
    \frac{\mathrm{d}^2p_{\rho}}{\mathrm{d}\Omega_0 \mathrm{d}\Omega_1} = 
    \mytr(\Psi_{E_1,\Omega_1}(\Psi_{E_0,\Omega_0}(\rho))) = 
    \frac{1}{\sigma_\mathrm{KN}(E_0) \sigma_\mathrm{KN}(E_1(E_0, \Omega_0))} \mytr(\Phi_{E_1,\Omega_1}(\Phi_{E_0,\Omega_0}(\rho))) \;.
\end{align}
Importantly, the map stays linear in the incoming polarization $\rho$.

By iterating this argument, the $N$-fold Compton scattering can thus be described by
\begin{align} \label{eq:post-scattering_state_unnormalized}
    \rho \longmapsto \rho' = \Psi_{E_{N-1}, \Omega_{N-1}}(\Psi_{E_{N-2},\Omega_{N-2}} (\dots (\Psi_{E_1,\Omega_1}(\Psi_{E_0,\Omega_0} ( \rho)))) ) \;,
\end{align}
with the probability density of scattering in $\Omega_0$, $\Omega_1$, ..., $\Omega_{N-1}$ given by
\begin{align}
    \frac{\mathrm{d}^N p_\rho}{\mathrm{d}\Omega_0 \mathrm{d}\Omega_1 \dots \mathrm{d}\Omega_{N-1}} = \mytr\left( \rho' \right) \;.
\end{align}
Integrating over finite angle intervals yields the scattering probability
\begin{align}
    p_\rho(s_0, s_1, ..., s_{N-1}) = \int_{s_0} \mathrm{d}\Omega_0 \int_{s_1} \mathrm{d}\Omega_1 ... \int_{s_{N-1}} \mathrm{d}\Omega_{N-1} \frac{\mathrm{d}^N p_\rho}{\mathrm{d}\Omega_0 \mathrm{d}\Omega_1 \dots \mathrm{d}\Omega_{N-1}} \;,
\end{align}
and the normalized post-scattering polarization state
\begin{align} \label{eq:post-scattering_state}
    \rho'_{s_0,s_1,...,s_{N-1}} = \frac{1}{p_\rho(s_0, s_1, ..., s_{N-1})} \int_{s_0} \mathrm{d}\Omega_0 \int_{s_1} \mathrm{d}\Omega_1 ... \int_{s_{N-1}} \mathrm{d}\Omega_{N-1} \rho' \;,
\end{align}
where $s_i \subseteq S^2$ for $i\in\{0,1,...,N-1\}$.
Note that \eqref{eq:post-scattering_state_unnormalized} and \eqref{eq:post-scattering_state} can not be computed using only the POVM density in \eqref{eq:POVM_density}.


\section{Expressing the fidelity bound with experimentally accessible ratios of coincidence counts}

Here we derive how to compute the fidelity lower bound in \eqref{eq:lower_bnd_fidelity_phi_minus} in the main text using only experimentally accessible quantities.

Using \eqref{eq:POVM_density} and \eqref{eq:rhogen} from the main text, we compute the probability densities
\begin{equation}
\begin{aligned} \label{eq:prob_density_AB_supp_mat}
    \frac{\mathrm{d}p_\rho}{\mathrm{d}\Omega_A \mathrm{d}\Omega_B} 
    &= \mytr((\Pi(\theta_A,\phi_A)\otimes \Pi(\theta_B,\phi_B)) \rho) \\
    &=\; G_A G_B S_{2,2} \sin (2 \phi_A) \sin (2 \phi_B)+ G_A G_B S_{1,1} \cos (2 \phi_A) \cos (2 \phi_B)+G_A G_B S_{1,2} \cos (2 \phi_A) \sin (2 \phi_B)+ \\
    &\qquad+G_A G_B S_{2,1} \sin (2 \phi_A) \cos (2 \phi_B)
    +G_A H_B S_{2,0} \sin (2 \phi_A)+G_A H_B S_{1,0} \cos (2 \phi_A)+\\
    &\qquad+G_B H_A S_{0,2} \sin (2 \phi_B)+G_B H_A S_{0,1} \cos (2 \phi_B)+H_A H_B  \;,
\end{aligned}
\end{equation}
where we already set $S_{0,0} = 1$ and defined $G_\ell := G(\theta_\ell)$ and $H_\ell := H(\theta_\ell)$ for $\ell\in\{A,B\}$.
Measuring the subsystems individually yields
\begin{align} \label{eq:prob_density_A_supp_mat}
    \frac{\mathrm{d}p_{\rho_A}}{\mathrm{d}\Omega_A} &= \mytr(\Pi(\theta_A,\phi_A) \rho_A) = H_A + G_A \cos(2\phi_A) S_{1,0} + G_A \sin(2\phi_A) S_{2,0} \;,\\
    \frac{\mathrm{d}p_{\rho_B}}{\mathrm{d}\Omega_B} &= \mytr(\Pi(\theta_B,\phi_B) \rho_B) = H_B + G_B \cos(2\phi_B) S_{0,1} + G_B \sin(2\phi_B) S_{0,2} \;. \label{eq:prob_density_B_supp_mat}
\end{align}
Following the discussion in the main text, we assume that detector counts are proportional to probabilities,
\begin{align}\label{eq:q_supp_mat}
    Q(\Tilde{\phi}_A, \Tilde{\phi}_B) &:= \mathcal{F}_{AB}
    \int_{\Tilde{\phi}_A -\Delta\phi_A/2}^{\Tilde{\phi}_A  +\Delta\phi_A/2} \mathrm{d}\phi_A
    \int_{\Tilde{\theta}_A-\Delta\theta_A/2}^{\Tilde{\theta}_A+\Delta\theta_A/2} \mathrm{d}\theta_A
    \int_{\Tilde{\phi}_B-\Delta\phi_B/2}^{\Tilde{\phi}_B+\Delta\phi_B/2} \mathrm{d}\phi_B
    \int_{\Tilde{\theta}_B-\Delta\theta_B/2}^{\Tilde{\theta}_B+\Delta\theta_B/2} \mathrm{d}\theta_B
    \sin\theta_A \sin\theta_B \; \frac{\mathrm{d}p_\rho}{\mathrm{d}\Omega_A \mathrm{d}\Omega_B} \;, \\
     \label{eq:q_i_supp_mat}
     Q_\ell  (\Tilde{\phi}_\ell) 
     &:= \mathcal{F}_\ell
     \int_{\Tilde{\phi}_\ell -\Delta\phi_\ell/2}^{\Tilde{\phi}_\ell  +\Delta\phi_\ell/2} \mathrm{d}\phi_\ell
     \int_{\Tilde{\theta}_\ell-\Delta\theta_\ell/2}^{\Tilde{\theta}_\ell+\Delta\theta_\ell/2} \mathrm{d}\theta_\ell
     \sin\theta_\ell \; \frac{\mathrm{d}p_{\rho_\ell}}{\mathrm{d}\Omega_\ell} \;,
\end{align}
where $\mathcal{F}_{AB}$ and $\mathcal{F}_\ell$ are the proportionality factors.
Because the only $\theta_\ell$-dependent quantities in \eqref{eq:prob_density_AB_supp_mat}-\eqref{eq:prob_density_B_supp_mat} are $G_\ell$ and $H_\ell$, and because these quantities appear linearly, we define
\begin{align}
    \mathcal{I}_{G_\ell} := \int_{\Tilde{\theta}_\ell-\Delta\theta_\ell/2}^{\Tilde{\theta}_\ell+\Delta\theta_\ell/2} \sin\theta_\ell \mathrm{d}\theta_\ell G(\theta_\ell) \;, \qquad
    \mathcal{I}_{H_\ell} := \int_{\Tilde{\theta}_\ell-\Delta\theta_\ell/2}^{\Tilde{\theta}_\ell+\Delta\theta_\ell/2} \sin\theta_\ell \mathrm{d}\theta_\ell H(\theta_\ell) \;, \label{eq:I_G_i_and_I_H_i_supp_mat}
\end{align}
and write
\begin{align}
    Q(\Tilde{\phi}_A, \Tilde{\phi}_B) &= \mathcal{F}_{AB}
    \int_{\Tilde{\phi}_A -\Delta\phi_A/2}^{\Tilde{\phi}_A  +\Delta\phi_A/2} \mathrm{d}\phi_A
    \int_{\Tilde{\phi}_B-\Delta\phi_B/2}^{\Tilde{\phi}_B+\Delta\phi_B/2} \mathrm{d}\phi_B \;
    \Big(\mathcal{I}_{G_A} \mathcal{I}_{G_B} S_{2,2} \sin (2 \phi_A) \sin (2 \phi_B)+ \notag\\
    &\qquad +
    \mathcal{I}_{G_A} \mathcal{I}_{G_B} S_{1,1} \cos (2 \phi_A) \cos (2 \phi_B)+
    \mathcal{I}_{G_A} \mathcal{I}_{G_B} S_{1,2} \cos (2 \phi_A) \sin (2 \phi_B)+ \notag\\
    &\qquad+
    \mathcal{I}_{G_A} \mathcal{I}_{G_B} S_{2,1} \sin (2 \phi_A) \cos (2 \phi_B)+
    \mathcal{I}_{G_A} \mathcal{I}_{H_B} S_{2,0} \sin (2 \phi_A)+
    \mathcal{I}_{G_A} \mathcal{I}_{H_B} S_{1,0} \cos (2 \phi_A)+\notag\\
    &\qquad+
    \mathcal{I}_{G_B} \mathcal{I}_{H_A} S_{0,2} \sin (2 \phi_B)+
    \mathcal{I}_{G_B} \mathcal{I}_{H_A} S_{0,1} \cos (2 \phi_B)+
    \mathcal{I}_{H_A} \mathcal{I}_{H_B} \Big) \;,
\end{align}
as well as
\begin{align}
    Q_A  (\Tilde{\phi}_A) 
     &= \mathcal{F}_A
     \int_{\Tilde{\phi}_A -\Delta\phi_A/2}^{\Tilde{\phi}_A  +\Delta\phi_A/2} \mathrm{d}\phi_A
     \big(\mathcal{I}_{H_A} + \mathcal{I}_{G_A} \cos(2\phi_A) S_{1,0} + \mathcal{I}_{G_A} \sin(2\phi_A) S_{2,0} \big) \;,\\
    Q_B  (\Tilde{\phi}_B) 
     &= \mathcal{F}_B
     \int_{\Tilde{\phi}_B -\Delta\phi_B/2}^{\Tilde{\phi}_B  +\Delta\phi_B/2} \mathrm{d}\phi_B
     \big(\mathcal{I}_{H_B} + \mathcal{I}_{G_B} \cos(2\phi_B) S_{0,1} + \mathcal{I}_{G_B} \sin(2\phi_B) S_{0,2} \big) \;.
\end{align}
The only integrals left to compute are of the form
\begin{align}
    \int_{\Tilde{\phi}_A -\Delta\phi_A/2}^{\Tilde{\phi}_A  +\Delta\phi_A/2} \mathrm{d}\phi_A
    \int_{\Tilde{\phi}_B-\Delta\phi_B/2}^{\Tilde{\phi}_B+\Delta\phi_B/2} \mathrm{d}\phi_B \cos(2\phi_A)^{c_A} \cos(2\phi_B)^{c_B} \sin(2\phi_A)^{s_A} \sin(2\phi_B)^{s_B} \;,
\end{align}
where $c_A, c_B, s_A, s_B \in\{0,1\}$, which can be evaluated analytically.
Doing so gives the ratios $\Bar{R}^{(1)}$, $\Bar{R}^{(2)}$, $\Bar{R}^{(1)}_A$, $\Bar{R}^{(2)}_B$, $\Bar{R}^{(1)}_B$, and $\Bar{R}^{(2)}_B$ in \eqref{eq:ratio1}-\eqref{eq:ratioB} as functions of $\tilde{\phi}_\ell$, $\Delta\phi_\ell$, $\tilde{\theta}_\ell$, $\Delta\theta_\ell$, and the density matrix elements $S_{1,1}$, $S_{2,2}$, $S_{0,1}$, $S_{0,2}$, $S_{1,0}$, and $S_{2,0}$.
Additionally, $\Bar{R}^{(1)}$ and $\Bar{R}^{(2)}$ depend on $S_{1,2}$ and $S_{2,1}$.

With this, we can evaluate \eqref{eq:ratioA}-\eqref{eq:ratioB} and solve them with respect to $S_{0,1}$, $S_{0,2}$, $S_{1,0}$, and $S_{2,0}$, giving
\begin{align}
    S_{0,1} &= \frac{\mathcal{I}_{H_B} \Delta\phi_B \left( (1-\Bar{R}^{(1)}_B\Bar{R}^{(2)}_B) \cos(2\delta\phi) - (1+\Bar{R}^{(1)}_B (\Bar{R}^{(2)}_B -2)) \sin(2\delta\phi) \right)}{\sin(\Delta\phi_B) \mathcal{I}_{G_B} (1+\Bar{R}^{(1)}_B \Bar{R}^{(2)}_B)} \;, \label{eq:S01_supp_mat}\\
    S_{0,2} &= \frac{\mathcal{I}_{H_B} \Delta\phi_B \left(  (\Bar{R}^{(1)}_B (2-\Bar{R}^{(2)}_B)-1) \cos(2\delta\phi)+ (\Bar{R}^{(1)}_B \Bar{R}^{(2)}_B -1) \sin(2\delta\phi)
     \right)}{\sin(\Delta\phi_B) \mathcal{I}_{G_B} (1+\Bar{R}^{(1)}_B \Bar{R}^{(2)}_B)} \;,\\
     S_{1,0} &= \frac{\mathcal{I}_{H_A} \Delta\phi_A \left( (1-\Bar{R}^{(1)}_A\Bar{R}^{(2)}_A) \cos(2\delta\phi) + (1+\Bar{R}^{(1)}_A (\Bar{R}^{(2)}_A -2)) \sin(2\delta\phi) \right)}{\sin(\Delta\phi_A) \mathcal{I}_{G_A} (1+\Bar{R}^{(1)}_A \Bar{R}^{(2)}_A)} \;,\\
    S_{2,0} &= \frac{\mathcal{I}_{H_A} \Delta\phi_A \left(  (\Bar{R}^{(1)}_A (2-\Bar{R}^{(2)}_A)-1) \cos(2\delta\phi)+ (1-\Bar{R}^{(1)}_A \Bar{R}^{(2)}_A ) \sin(2\delta\phi)
     \right)}{\sin(\Delta\phi_A) \mathcal{I}_{G_A} (1+\Bar{R}^{(1)}_A \Bar{R}^{(2)}_A)} \;. \label{eq:S_20_supp_mat}
\end{align}
Furthermore, evaluating \eqref{eq:ratio1} and \eqref{eq:ratio2}, and solving for $S_{1,1}$ and $S_{2,2}$ gives
\begin{align} \label{eq:S11_supp_mat}
    S_{1,1} &= \frac{-\cos^2(2\delta\phi) a + \sin^2(2\delta\phi) b + \sin(4\delta\phi) \frac{S_{1,2}-S_{2,1}}{2}}{\cos(4\delta\phi)} \;,\\
    S_{2,2} &= \frac{-\sin^2(2\delta\phi) a + \cos^2(2\delta\phi) b + \sin(4\delta\phi) \frac{S_{1,2}-S_{2,1}}{2}}{\cos(4\delta\phi)} \;, \label{eq:S22_supp_mat}
\end{align}
where
\begin{equation}
\begin{aligned}
    a &= \frac{f(\Bar{R}^{(1)})\big(u( \cos(2\delta\phi) S_{1,0} + \sin(2\delta\phi) S_{2,0}) + w\big) + v (\cos(2\delta\phi) S_{0,1}-\sin(2\delta\phi) S_{0,2})}{\mathcal{I}_{G_A} \mathcal{I}_{G_B} \sin(\Delta\phi_A) \sin(\Delta\phi_B)} \;, \\
    b &= \frac{f(\Bar{R}^{(2)})\big(u( \sin(2\delta\phi) S_{1,0} - \cos(2\delta\phi) S_{2,0}) - w\big) - v (\sin(2\delta\phi) S_{0,1}+\cos(2\delta\phi) S_{0,2})}{\mathcal{I}_{G_A} \mathcal{I}_{G_B} \sin(\Delta\phi_A) \sin(\Delta\phi_B)} \;,
\end{aligned}
\end{equation}
with $f(t) = \frac{t-1}{t+1}$, and $u = \mathcal{I}_{G_A} \mathcal{I}_{H_B} \Delta\phi_B \sin(\Delta\phi_A)$, $v = \mathcal{I}_{G_B} \mathcal{I}_{H_A} \Delta\phi_A \sin(\Delta\phi_B)$, and $w = \mathcal{I}_{H_A} \mathcal{I}_{H_B} \Delta\phi_A \Delta\phi_B$.

Substituting \eqref{eq:S11_supp_mat} and \eqref{eq:S22_supp_mat} into the fidelity lower bound $F_{\Phi^-} \geq \frac{S_{2,2}- S_{1,1}}{2}$ in \eqref{eq:lower_bnd_fidelity_phi_minus} in the end matter of the main text, and subsequently substituting \eqref{eq:S01_supp_mat}-\eqref{eq:S_20_supp_mat}, we obtain
\begin{align} \label{eq:lower_bnd_with_ratios_supp_mat_final}
    F_{\Phi^-} \geq\frac{S_{2,2}- S_{1,1}}{2} = \mathcal{I}_0
    \left(
    \frac{f(\Bar{\mathcal{R}}^{(1)})-\Bar{\mathcal{R}}_A^{(1)} f(\Bar{\mathcal{R}}^{(2)})}{1+\Bar{\mathcal{R}}_A^{(1)} \Bar{\mathcal{R}}_A^{(2)}} + \frac{1-\Bar{\mathcal{R}}_B^{(1)}}{1+\Bar{\mathcal{R}}_B^{(1)} \Bar{\mathcal{R}}_B^{(2)}}
    \right) \;,
\end{align}
where
\begin{align}
    \mathcal{I}_0 = \frac{\Delta\phi_A \Delta\phi_B}{\sin\Delta\phi_A\sin\Delta\phi_B}\frac{\mathcal{I}_{H_A} \mathcal{I}_{H_B}}{\mathcal{I}_{G_A} \mathcal{I}_{G_B}} \;.
\end{align}
Note that the dependency of $S_{1,2}$ and $S_{2,1}$ in \eqref{eq:S11_supp_mat} and \eqref{eq:S22_supp_mat} is such that it cancels out in \eqref{eq:lower_bnd_with_ratios_supp_mat_final}.


\section{Experimental proposal and rough error estimate}

In this section, we present some numerical values for a potential experimental realization and compute a rough error estimate.
This shows that it is principally possible to certify entanglement in current experiments.
In particular, we emphasize that many such experiments have been done (cf. Ref.~\cite{Watts_Photonquantumentanglement_2021, Bordes_FirstDetailedStudy_2024, Moskal_Nonmaximalentanglementphotons_2025, Ivashkin_Testingentanglementannihilation_2023, Bala_Probingarbitrarypolarized_2025}) and that one can potentially compute the value of our observable \eqref{eq:rho22-rho11_using_ratios} by analyzing existing data.

First, note that $\tilde{\phi}_A$ and $\tilde{\phi}_B$ are variables for the different observable ratios in \eqref{eq:ratio1}-\eqref{eq:ratioB}.
For the polar and azimuthal specifications we propose to use $\tilde{\theta}_A = \tilde{\theta}_B = \tilde{\theta}_0 = 82^\circ$ with $\Delta\theta = 20^\circ$, and $\Delta\phi = \pi/4$, respectively.
The reasons are the following:
Many known results focus on the polar region around $\approx 81.7^\circ$, where the so-called visibility is maximized.
Consulting Fig.~10 in Ref.~\cite{Bala_Probingarbitrarypolarized_2025} also shows that the obtained counts are largely centered around this area.
This increases statistics and reduces measurement time.
We also choose a large azimuthal resolution of $\Delta\phi = \pi/4$ for this last reason, even though better resolutions are available (cf. Ref.~\cite{Watts_Photonquantumentanglement_2021, Bordes_FirstDetailedStudy_2024,  Moskal_Nonmaximalentanglementphotons_2025, Ivashkin_Testingentanglementannihilation_2023, Bala_Probingarbitrarypolarized_2025}).

For the errors per measured ratio \eqref{eq:ratio1}-\eqref{eq:ratioB}, we extrapolate from the errors reported in the aforementioned publications.
They computed similar ratios and obtained relative errors between 0.74\% \cite{Ivashkin_Testingentanglementannihilation_2023} and 2.2\% \cite{Watts_Photonquantumentanglement_2021}.
Here, we estimate the relative error very conservatively at 5\%.
For simplicity, we furthermore assume that the errors are independent and uncorrelated.
We admit that this assumption is not valid in realistic experiments, but it is helpful for a first, rough error estimate.
A more detailed propagation of errors is deferred to future work.

Under these specifications and assumptions, we can write $\Delta_R = 0.05 \, |R|$ for the uncertainties of the ratios $R\in \boldsymbol{R} =\{\Bar{R}^{(1)}, \Bar{R}^{(2)}, \Bar{R}^{(1)}_A, \Bar{R}^{(2)}_B, \Bar{R}^{(1)}_B, \Bar{R}^{(2)}_B\}$.
Denoting the lower bound in \eqref{eq:rho22-rho11_using_ratios} by $\mathcal{L}$, its absolute uncertainty is then given by
\begin{align}
    \Delta_\mathcal{L} = 
    \sqrt{\sum_{R \in \boldsymbol{R}} \left( \frac{\partial \mathcal{L}}{\partial R} \, \Delta_R \right)^2} = 
    0.05 \; \sqrt{ \sum_{R \in \boldsymbol{R}} \left( \frac{\partial \mathcal{L}}{\partial R} \, |R| \right)^2} \;.
\end{align}
The relative uncertainty is then
\begin{align}
    \delta_\mathcal{L} = \frac{\Delta_\mathcal{L}}{ \mathcal{L}} \;.
\end{align}
Inserting the angle specifications above, we obtain an absolute uncertainty of
\begin{align}
    \Delta_\mathcal{L} \approx 0.083 \,\hat{=}\, 8.3\%\;.
\end{align}
If the initial polarization state is the maximally entangled Bell state $|\Phi^-\rangle = \frac{1}{\sqrt{2}}(|R\rangle\otimes |R\rangle - |L\rangle\otimes |L\rangle)$, then $\mathcal{L} = 1$ and thus the relative error is also $\delta_\mathcal{L} \approx 0.083 \,\hat{=}\, 8.3\%$.
As the state is certified as entangled for $1/2<\mathcal{L}\leq1$, this accuracy is enough to witness entanglement.
Furthermore, separable states satisfy $\max_{\rho\in\mathrm{SEP}}\mathcal{L} = 1/2$, and thus the scheme reliably distinguishes between $|\Phi^-\rangle$ and any separable state


\end{document}